\documentclass[aps,amsmath,amssymb,prb,twocolumn,superscriptaddress,floatfix]{revtex4-2}

\usepackage{graphicx}
\usepackage[svgnames]{xcolor}
\usepackage{hyperref}

\newcommand{\muB}{\mu_{\text{B}}}
\newcommand{\Tc}{T_{\text{c}}}

\begin{document}

\title{Uncertainties in experiments on strongly-coupled vacuum field modification of superconductivity and ferromagnetism in solids}

\author{J. R. Cooper}
\affiliation{Cavendish Laboratory, University of Cambridge, J. J. Thomson Avenue, Cambridge, CB3 0HE, United Kingdom}

\author{L. Forr\'o}
\affiliation{Stavropoulos Center for Complex Quantum Matter, Department of Physics and Astronomy, University of Notre Dame, Notre Dame, Indiana 46556, United States; e-mail: lforro@nd.edu}

\author{A. J\'anossy}
\affiliation{Institute of Physics, Budapest University of Technology and Economics, 1111 Budapest, Hungary}

\date{\today}

\begin{abstract}
We discuss recent experiments in which fine particles of the organic superconductor Rb$_3$C$_{60}$ or the cuprate superconductor YBa$_2$Cu$_3$O$_{6+x}$ are held in a polystyrene film that is spin-coated on to a silicon substrate with or without an intervening gold, or another inert metallic layer. From SQUID magnetisation data for Rb$_3$C$_{60}$ there appears to be a striking and completely unexpected increase in the superconducting transition temperature from $30$ to $45$~K, which is ascribed to coupling between the electrons in the superconductor and vacuum fluctuations in the electromagnetic field just above the metallic film. We argue that this could be a non-intrinsic effect associated with the presence of solid oxygen in the Pyrex sample tube. We suggest that the ferromagnetic SQUID signal observed for YBa$_2$Cu$_3$O$_{6+x}$ particles in polystyrene could be attributed to ferromagnetic particles or magnetic clusters of unknown origin.
\end{abstract}

\maketitle
\section{Introduction}

For over a decade there has been much work on how strong coupling to light influences material properties and chemical reactions \cite{reviewpaper1,reviewpaper2,reviewpaper3}. In most of the experiments the materials are placed in cavities or other resonators. The cavity increases the light intensity from an external source at certain optimal frequencies needed to reach the light-matter strong coupling regime.

Here we focus on more recent claims that the electronic properties of some superconducting compounds can be strongly altered by coupling to vacuum field fluctuations near suitable polarizable materials \cite{archivepaper,FerroYBCO,ScienceRev}. These papers report a substantial increase of the superconducting transition temperature ($\Tc$) of Rb$_3$C$_{60}$ and the appearance of a previously unknown ferromagnetic phase persisting even above ambient temperature in the superconductor YBa$_2$Cu$_3$O$_{6+x}$~(YBCO$_{6+x}$). In order to observe these extraordinary changes, the materials were embedded in polymeric polystyrene~(PS) and spin-coated on a gold (Au) layer. No external light source was applied to induce the large changes in electronic properties claimed. Earlier measurements of the infrared spectrum using  resonators to enhance light intensity established that PS infrared vibrations can strongly couple to light from an external source. The new experiments in Refs. \onlinecite{archivepaper,FerroYBCO} show a slight difference in the coupling to an external light source of PS with or without embedded superconductors. The claims of large changes in the electronic properties of the superconductors attributed to vacuum field fluctuations are based on magnetometry of samples layered on to highly conducting Au or insulating silicon in the absence of any external light. 

The electronic properties of both superconductors in these studies have been thoroughly investigated since they were discovered more than three decades ago~\cite{sconref1,sconref2,Rb3C60scon1,Rb3C60scon2,Rb3C60scon3}. Thus reports of unexpected and unexplained large changes in their physical properties should be verified with utmost care. The works \cite{archivepaper,FerroYBCO} have been cited in over a hundred, mainly theoretical papers, but the primary article on superconductivity enhancement in Rb$_3$C$_{60}$ has only appeared on the cond-mat archive~\cite{archivepaper} that is cited in a review~\cite{ScienceRev} but not in a regular refereed scientific journal. Here we wish to draw attention to possible uncertainties in the assessment of the experiments, which may be useful for any researchers, particularly younger people, who are tasked with verifying the experimental data reported for Rb$_3$C$_{60}$ and YBCO$_{6+x}$. We detail difficulties encountered in measuring the magnetization of small samples of chemically reactive materials. We do not comment on aspects of vacuum fluctuations that are not directly related to superconductivity, for example the significance of the optical measurements on the PS films~\cite{archivepaper} or low temperature quantum Hall experiments~\cite{Hallpaper} or the effect of high intensity light on superconductivity~\cite{pumppaper}.

\begin{figure}[h!tb]
      \includegraphics[width=\columnwidth]{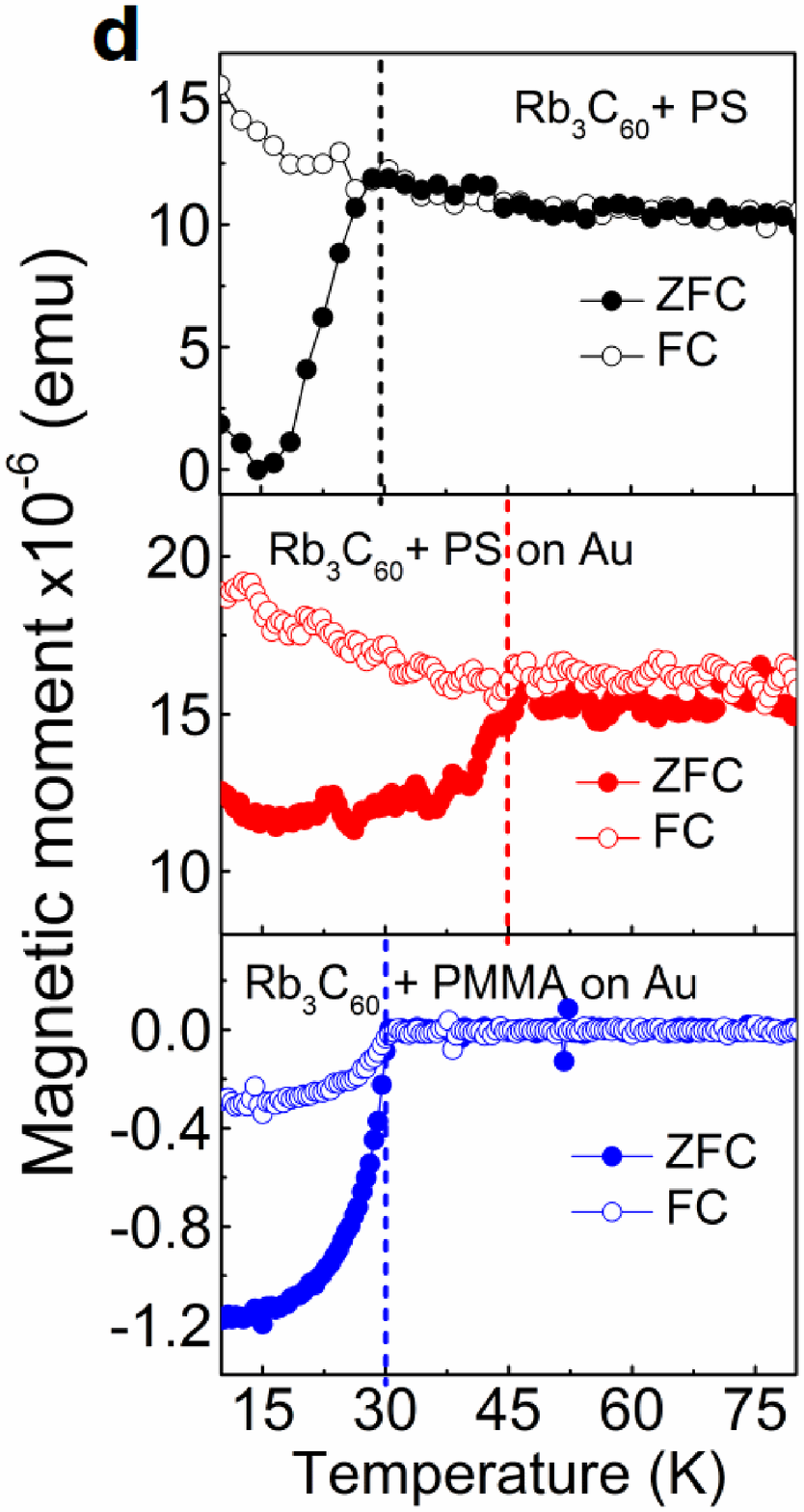}
      \caption{Part of Fig. 3 of Ref.~\onlinecite{archivepaper} showing proposed evidence for the enhancement of the superconducting transition temperature of Rb$_3$C$_{60}$ from $30$ to $45$~K by vacuum fluctuations associated with a Au layer underneath the polystyrene~(PS) film containing particles of Rb$_3$C$_{60}$, upper figure: sample attached to a Si plate. middle: sample attached to thin Au layer on Si. lower: sample embedded in a different polymer (PMMA). The sample temperatures are cycled between high temperatures and $4$~K. All measurements are made while heating the sample from $4$~K in a magnetic field of $100$~G. ZFC: zero field cooling. FC: $100$~G field cooled.}
      \label{Fig1}
\end{figure}

\section{Uncertainity in assessment of experiments on enhancement of superconductivity in \texorpdfstring{$\mathrm{Rb}_3\mathrm{C}_{60}$}{Lg}}

\begin{figure}[h!tb]
    \includegraphics[width=\columnwidth]{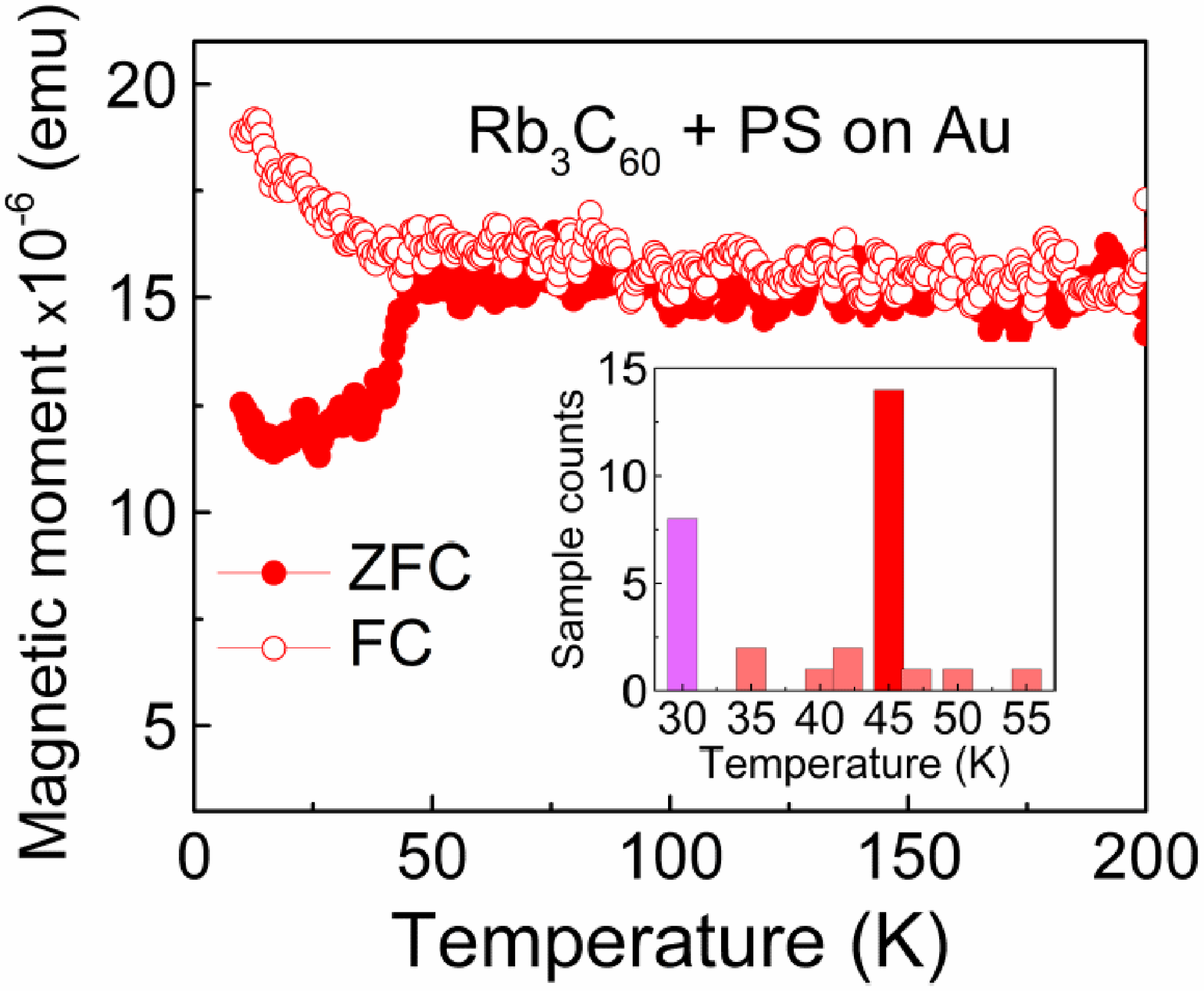}
    \caption{Part of Fig. 3 of Ref. \onlinecite{archivepaper} showing statistics of the $\Tc$ values observed.}
    \label{Fig2}
\end{figure}

The superconducting transition temperature was determined from the temperature dependence of the magnetic moment measured by SQUID magnetometry. All measurements were performed while heating the sample from low temperatures in a $100$~G magnetic field after cooling in $100$~G or in zero field. Below a well-defined temperature, defined as the superconducting~(s/c) transition temperature, a difference was observed between the zero field~(ZFC) and the $100$ G~($10$ mT) field-cooled~(FC) data. The difference is attributed to hysteresis in the Meissner effect. This kind of hysteresis is well-established in bulk samples of Rb$_3$C$_{60}$ and many other superconductors. Without going into detail, we can say that the magnitudes of the signal in the two top panels of Fig.~\ref{Fig1} are approximately consistent with expectations for  $6$~wt\% of spherical particles of Rb$_3$C$_{60}$, diameter $1~\mu$m in a $4\times 4$~mm$^2$, $4~\mu$m thick PS film~\cite{archivepaper} after taking into account the London penetration depth of $0.1~\mu$m for Rb$_3$C$_{60}$~\cite{McHenry}. However, there is uncertainty in that the density of the PS film and the amount of Rb$_3$C$_{60}$ actually in the film after spin coating are not known. Note that the magnetic susceptibility, $\sim-1/(4\pi)$ emu/cm$^3$, in the Meissner state is very large compared with the usual paramagnetic or diamagnetic response of solids. The smaller signal for the sample in the polymer PMMA than for PS suggests that in these experiments the s/c fraction, i.e., the amount of Rb$_3$C$_{60}$ in the film, was not well-controlled, as also suggested by the insert to Fig.~\ref{Fig2}.

One uncertainty with this approach is that the transition temperature may signal a hysteretic magnetic transition from an extraneous material in the sample holder or the sample. Some possibilities  considered below are: (i) magnetic impurities in the Pyrex tube material. (ii) magnetic phases of O$_2$ adsorbed on the inner wall of the Pyrex tube or in the porous PS film. (iii) thermometry. (iv) magnetic phases of Rb$_1$C$_{60}$.

\section{Problems with the Pyrex tube}

The samples are held in a Pyrex tube which contains a small amount of magnetic impurities,~\cite{Hurd1963} giving a $T$-dependent background signal. Details of the empty holder signal and any subtraction procedure were not published. The available susceptibility data for Pyrex~\cite{Hurd1963} can be fitted to an $A + C/(T+\Theta)$ law where $A$ is a constant, $C=1.3\times 10^{-5}$~emu$\cdot$K/g Pyrex and $\Theta=2.8$~K. $\Theta$ is small and this value of $C$ corresponds to $0.05$~wt\% of Fe impurities with an effective moment of $5.9 \muB$. So impurities in Pyrex are unlikely to order magnetically at the temperatures measured. However, it results in a temperature dependent background that is large compared to the sample signal. The background can be positive or negative depending on whether there is more Pyrex in the central measuring coil of the SQUID magnetometer or in the two outer coils, which can cause complications when centering the sample.

\section{Possibility of \texorpdfstring{$\mathrm{O}_2$}{Lg} contamination}

\begin{figure}
    \includegraphics[width=\columnwidth]{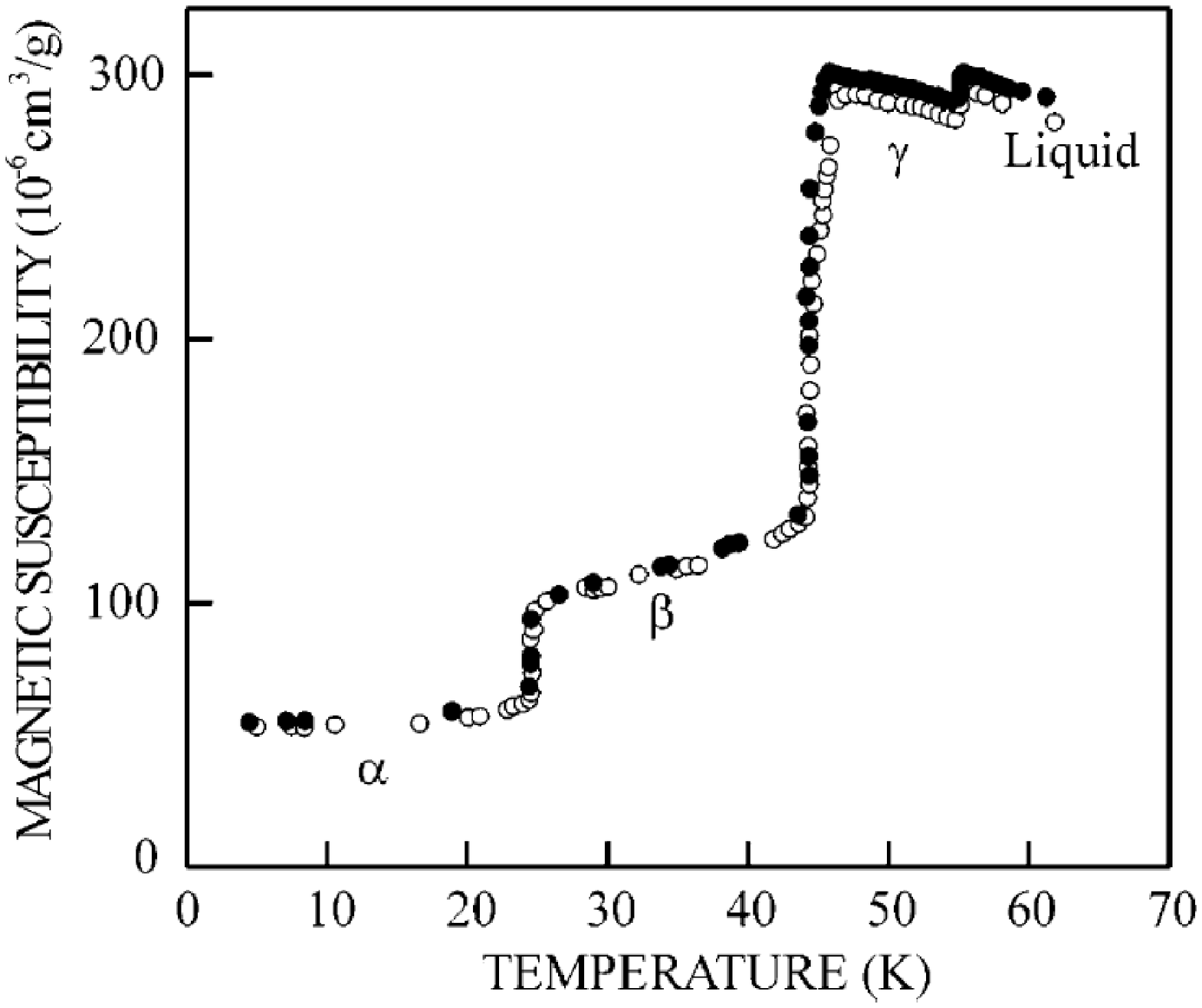}
    \caption{The magnetic susceptibility of condensed oxygen from Ref.~\onlinecite{Freiman2004}.}
    \label{oxygen}
\end{figure}

The susceptibility of condensed oxygen is shown in Fig. \ref{oxygen}~\cite{Freiman2004}. Out of the three solid phases, $\alpha$ and $\beta$ are antiferromagnetic while $\gamma$ is paramagnetic. At the $45$ K transition from the antiferromagnetic $\beta$ phase to the paramagnetic $\gamma$ phase there is a large jump in the magnetic susceptibility of $1.7 \times 10^{-4}$~emu/g, that is a change in magnetic moment of $1.7 \times 10^{-2}$~emu/g at $100$~G. In order to account for the signal in Fig.~\ref{Fig1} there would need to be $10^{-5}/1.7 \times 10^{-2}\simeq 0.5$~mg O$_2$ condensed inside the sample tube.

There is a volume change at the $\gamma-\beta$ phase transition~\cite{Freiman2004} so this is clearly first order and is therefore prone to supercooling or superheating effects giving hysteresis. There is also short-range magnetic order in the $\beta$-phase, which by analogy with data for spin glasses, could be very sensitive to low magnetic fields. Note that the signal would change if the tube were heated above $60$~K because in that case the oxygen could be re-condensed in a different part of the tube on cooling again. Depending on the details of the cooling protocol, this could also cause hysteresis in the data.

We note that $0.5$~mg of solid oxygen has a volume of $0.5/1.24$~mm$^3$ i.e. for an area of $4\times 4$~mm$^2$, this corresponds to a thickness of $25~\mu$m, so the $4~\mu$m thick PS film would have to be extremely porous to accommodate it. In the experiments reported in Ref. \onlinecite{archivepaper} several samples have a hysteretic transition temperature of $45$ K ascribed to the onset of superconductivity. The authors have taken precautionary measures to avoid O$_2$ contamination during sample preparation and while placing samples in the magnetometer. However, further confirmation that this does not affect the results is desirable. A possible test might be to determine the amount of oxygen gas within the tube after measurement in the SQUID magnetometer. AC susceptibility measurements using small detection coils or a tunnel diode oscillator~\cite{Gianetta2022} would help avoid centering problems in the magnetometer.

\section{Thermometry}

During the heating cycle, the sample temperature may lag behind the temperature measured outside the Pyrex sample holder unless there is some He gas inside the Pyrex tube to ensure good thermal contact. It may be worth noting that at room temperature Pyrex is permeable to helium gas. So if the sample holder is inserted into the SQUID magnetometer at room temperature and held there for a few minutes, there will be helium gas around the sample giving good thermal contact to the surrounding isothermal chamber, but if it were inserted at a lower temperature, which is sometimes done to save time, this would not be the case.

\section{Other possibilities}

\subsection{SDW transition in orthorhombic \texorpdfstring{(o)-$\mathrm{RbC}_{60}$}{Lg} below about \texorpdfstring{$50~\mathrm{K}$}{Lg}}

\begin{figure}
    \includegraphics[width=\columnwidth, height=150mm]{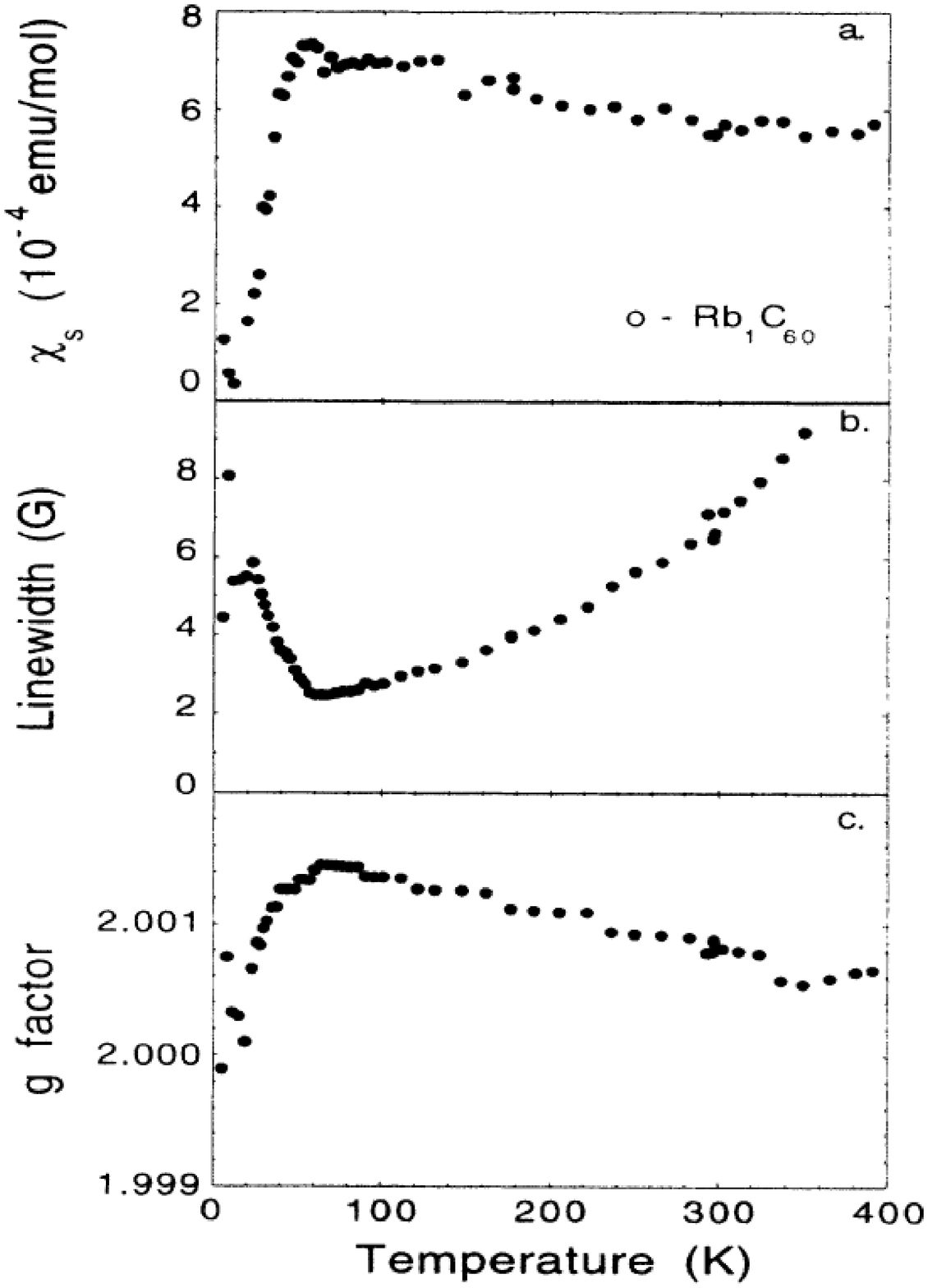}
    \caption{Fig. 2 of Ref. \onlinecite{RbC60SDW1994PRL}. Spin susceptibility, ESR linewidth and $g$-factor of orthorhombic RbC$_{60}$.}
    \label{SDW}
\end{figure}

Some of the Rb$_3$C$_{60}$ superconductor can be lost or converted to a Rb compound with an altered stoichiometry during sample preparation. Among the Rb doped C$_{60}$ compounds, RbC$_{60}$ is stable in air and enters an antiferromagnetic spin density wave (SDW) state below about $50$~K~\cite{RbC60SDW1994PRL}. It is difficult to synthesize a phase pure sample of Rb$_3$C$_{60}$ without at least some traces of RbC$_{60}$ in the batch. One might lose superconductivity if for some reason Rb$_3$C$_{60}$ is oxidized, or if the Rb reacts with the metallic layer. But the change in spin susceptibility as the SDW develops in (o)-RbC$_{60}$ is far too small to account for the anomaly in the middle panel of Fig.~\ref{Fig1}. The top panel of Fig.~\ref{SDW} for (o)-RbC$_{60}$ shows a decrease in spin susceptibility of $7 \times 10^{-4}$~emu/mol at $45$ K caused by the SDW~\cite{RbC60SDW1994PRL}. The X-ray density of (o)-RbC$_{60}$ is $2.03$~g/cm$^3$ so for a molecular weight of $805$ this would give a SQUID magnetometer signal of $1.76 \times 10^{-4}$~emu for $1$~cm$^3$ at $100$~G. A typical s/c signal (say $1/10$ of full diamagnetism) is $-10/(4\pi)$ or $-0.8$~emu for $1$~cm$^3$ at $100$~G, a factor of $5 \times 10^3$ larger than the SDW signal from RbC$_{60}$.

So the only effect of the decomposition of Rb$_3$C$_{60}$ into RbC$_{60}$ is to reduce the superconducting fraction, i.e., the amount of Rb$_3$C$_{60}$ in the film.

\subsection{Possible effect of stress on \texorpdfstring{$\mathrm{Rb}_3\mathrm{C}_{60}$}{Lg}}

Another possibility is that thermal contraction of the polystyrene exerts unusual non-hydrostatic pressure on Rb$_3$C$_{60}$, that gives the increase in the s/c $\Tc$ observed. This would be a surprising and interesting result, but is unlikely because hydrostatic pressure reduces $\Tc$ of Rb$_3$C$_{60}$ by approximately $1$~K/kilobar~\cite{Schilling}.

\section{Conclusions for \texorpdfstring{$\mathrm{Rb}_3\mathrm{C}_{60}$}{Lg}}

It seems that unwanted oxygen contamination is a possible non-intrinsic explanation of the results shown. A rather large amount is needed, perhaps arising from outgassing of the Pyrex tube when it was melted at one end and sealed. Other possibilities are less likely. \vspace{5mm}

\section{Uncertainty in the assessment of experiments on the enhancement of ferromagnetism in \texorpdfstring{$\mathrm{YBCO}_{6+x}$}{Lg}}

Ref.~\cite{FerroYBCO} reported a large enhancement of the ferromagnetism of YBa$_2$Cu$_3$O$_{6+x}$ nanoparticles under a cooperative strong coupling. According to the report, the saturation magnetic moment reaches $0.9\muB$ per mole YBa$_2$Cu$_3$O$_{6+x}$ and competes with superconductivity at low temperatures. However, for YBCO powder, YBCO embedded in PS and layered on Si or embedded in PMMA polymer and layered on Au only a negligibly small room temperature ferromagnetism was observed.

There are often experimental problems in establishing the presence of small ferromagnetic signals using SQUID magnetometry and showing that they are intrinsic~\cite{Garcia2009}. It is also known that grinding YBCO$_{7}$ samples in air and to a lesser extent under argon~\cite{Panagopoulos1996} produces a magnetic surface layer which can be only repaired by using a specific high $T$ annealing process. One of the compositions at the surface observed by high resolution electron microscopy in Ref.~\onlinecite{Panagopoulos1996} was Cu$_8$O.

An intriguing speculation is that this compound is in fact super-paramagnetic or even ferromagnetic. The X-ray diffraction data in the supplementary material of Ref. \onlinecite{FerroYBCO} show the presence of several other phases particularly the green phase Y$_2$BaCuO$_5$ and also BaCuO$_{2+x}$ which has a complicated crystal structure~\cite{Guskos1995} with CuO polyhedra (clusters) that are expected~\cite{Guskos1995} to exhibit ferromagnetic intra-cluster interactions. The signal observed in~Ref.~\onlinecite{FerroYBCO} is relatively large. 

The apparent ferromagnetism at room temperature could be caused  by magnetic  clusters in BaCuO$_{2}$ giving curvature in magnetisation-field plots at low fields. These clusters can be stabilised
by the inclusion of carbon~\cite{Aranda}. One possible hypothesis is that they would be different when the films are deposited on a silicon substrate which has an oxide layer rather than on a Au layer. Such clusters would have to be unusually large in order to give significant  non-linearity at $1000$ G ($0.1$T) and $300$K.

Another complicating factor is that the superconducting $\Tc$ of YBCO$_{6+x}$ is very dependent on the oxygen content of the CuO chains and on the degree of oxygen order in these chains which in turn depends on the cooling rate used below $400$~K~\cite{Monod2012}. Oxygen disorder in the CuO chains gives a Curie term in the spin susceptibility which depends both on the cooling rate~\cite{Monod2012} and on how long the sample has been held at room temperature~\cite{Kokanovic2016}.

Finally we note that if the experiments are repeated it would be wise to use non-magnetic copper bronze tools throughout rather than stainless steel, and to start with phase pure superconducting YBCO to reduce the probability of extrinsic super-paramagnetic or ferromagnetic phases. 

\section{Conclusion for \texorpdfstring{$\mathrm{YBCO}_{6+x}$}{Lg}}

There is a strong possibility that the effect observed is not caused by vacuum fluctuations coupled to the YBCO$_{6+x}$ superconductor. It may rather arise from as yet undetermined, ferromagnetic foreign particles, phase or magnetic clusters such as those present in BaCuO$_{2+x}$.

\section{Acknowledgements}
We are indebted to Dr. Bence G. M\'arkus for editing this work.

\bibliography{Vacuum_SC_comment}

%apsrev4-2.bst 2019-01-14 (MD) hand-edited version of apsrev4-1.bst
%Control: key (0)
%Control: author (72) initials jnrlst
%Control: editor formatted (1) identically to author
%Control: production of article title (-1) disabled
%Control: page (0) single
%Control: year (1) truncated
%Control: production of eprint (0) enabled
\begin{thebibliography}{25}%
\makeatletter
\providecommand \@ifxundefined [1]{%
 \@ifx{#1\undefined}
}%
\providecommand \@ifnum [1]{%
 \ifnum #1\expandafter \@firstoftwo
 \else \expandafter \@secondoftwo
 \fi
}%
\providecommand \@ifx [1]{%
 \ifx #1\expandafter \@firstoftwo
 \else \expandafter \@secondoftwo
 \fi
}%
\providecommand \natexlab [1]{#1}%
\providecommand \enquote  [1]{``#1''}%
\providecommand \bibnamefont  [1]{#1}%
\providecommand \bibfnamefont [1]{#1}%
\providecommand \citenamefont [1]{#1}%
\providecommand \href@noop [0]{\@secondoftwo}%
\providecommand \href [0]{\begingroup \@sanitize@url \@href}%
\providecommand \@href[1]{\@@startlink{#1}\@@href}%
\providecommand \@@href[1]{\endgroup#1\@@endlink}%
\providecommand \@sanitize@url [0]{\catcode `\\12\catcode `\$12\catcode
  `\&12\catcode `\#12\catcode `\^12\catcode `\_12\catcode `\%12\relax}%
\providecommand \@@startlink[1]{}%
\providecommand \@@endlink[0]{}%
\providecommand \url  [0]{\begingroup\@sanitize@url \@url }%
\providecommand \@url [1]{\endgroup\@href {#1}{\urlprefix }}%
\providecommand \urlprefix  [0]{URL }%
\providecommand \Eprint [0]{\href }%
\providecommand \doibase [0]{https://doi.org/}%
\providecommand \selectlanguage [0]{\@gobble}%
\providecommand \bibinfo  [0]{\@secondoftwo}%
\providecommand \bibfield  [0]{\@secondoftwo}%
\providecommand \translation [1]{[#1]}%
\providecommand \BibitemOpen [0]{}%
\providecommand \bibitemStop [0]{}%
\providecommand \bibitemNoStop [0]{.\EOS\space}%
\providecommand \EOS [0]{\spacefactor3000\relax}%
\providecommand \BibitemShut  [1]{\csname bibitem#1\endcsname}%
\let\auto@bib@innerbib\@empty
%</preamble>
\bibitem [{\citenamefont {Carusotto}\ and\ \citenamefont
  {Ciuti}(2013)}]{reviewpaper1}%
  \BibitemOpen
  \bibfield  {author} {\bibinfo {author} {\bibfnamefont {I.}~\bibnamefont
  {Carusotto}}\ and\ \bibinfo {author} {\bibfnamefont {C.}~\bibnamefont
  {Ciuti}},\ }\href@noop {} {\bibfield  {journal} {\bibinfo  {journal} {Rev.
  Mod. Phys.}\ }\textbf {\bibinfo {volume} {85}},\ \bibinfo {pages} {299}
  (\bibinfo {year} {2013})}\BibitemShut {NoStop}%
\bibitem [{\citenamefont {T\"orm\"a}\ and\ \citenamefont
  {Barnes}(2014)}]{reviewpaper2}%
  \BibitemOpen
  \bibfield  {author} {\bibinfo {author} {\bibfnamefont {P.}~\bibnamefont
  {T\"orm\"a}}\ and\ \bibinfo {author} {\bibfnamefont {W.~L.}\ \bibnamefont
  {Barnes}},\ }\href@noop {} {\bibfield  {journal} {\bibinfo  {journal}
  {Reports on Progress in Physics}\ }\textbf {\bibinfo {volume} {78}},\
  \bibinfo {pages} {013901} (\bibinfo {year} {2014})}\BibitemShut {NoStop}%
\bibitem [{\citenamefont {Ruggenthaler}\ \emph {et~al.}(2018)\citenamefont
  {Ruggenthaler}, \citenamefont {Tancogne-Dejean}, \citenamefont {Flick},
  \citenamefont {Appel},\ and\ \citenamefont {Rubio}}]{reviewpaper3}%
  \BibitemOpen
  \bibfield  {author} {\bibinfo {author} {\bibfnamefont {M.}~\bibnamefont
  {Ruggenthaler}}, \bibinfo {author} {\bibfnamefont {N.}~\bibnamefont
  {Tancogne-Dejean}}, \bibinfo {author} {\bibfnamefont {J.}~\bibnamefont
  {Flick}}, \bibinfo {author} {\bibfnamefont {H.}~\bibnamefont {Appel}},\ and\
  \bibinfo {author} {\bibfnamefont {A.}~\bibnamefont {Rubio}},\ }\href@noop {}
  {\bibfield  {journal} {\bibinfo  {journal} {Nat. Rev. Chem.}\ }\textbf
  {\bibinfo {volume} {2}},\ \bibinfo {pages} {0118} (\bibinfo {year}
  {2018})}\BibitemShut {NoStop}%
\bibitem [{\citenamefont {Thomas}\ \emph {et~al.}(2019)\citenamefont {Thomas},
  \citenamefont {Devaux}, \citenamefont {Nagarajan}, \citenamefont {Chervy},
  \citenamefont {Seidel}, \citenamefont {Hagenm\"uller}, \citenamefont
  {Sch\"utz}, \citenamefont {Schachenmayer}, \citenamefont {Genet},
  \citenamefont {Pupillo},\ and\ \citenamefont {Ebbesen}}]{archivepaper}%
  \BibitemOpen
  \bibfield  {author} {\bibinfo {author} {\bibfnamefont {A.}~\bibnamefont
  {Thomas}}, \bibinfo {author} {\bibfnamefont {E.}~\bibnamefont {Devaux}},
  \bibinfo {author} {\bibfnamefont {K.}~\bibnamefont {Nagarajan}}, \bibinfo
  {author} {\bibfnamefont {T.}~\bibnamefont {Chervy}}, \bibinfo {author}
  {\bibfnamefont {M.}~\bibnamefont {Seidel}}, \bibinfo {author} {\bibfnamefont
  {D.}~\bibnamefont {Hagenm\"uller}}, \bibinfo {author} {\bibfnamefont
  {S.}~\bibnamefont {Sch\"utz}}, \bibinfo {author} {\bibfnamefont
  {J.}~\bibnamefont {Schachenmayer}}, \bibinfo {author} {\bibfnamefont
  {C.}~\bibnamefont {Genet}}, \bibinfo {author} {\bibfnamefont
  {G.}~\bibnamefont {Pupillo}},\ and\ \bibinfo {author} {\bibfnamefont {T.~W.}\
  \bibnamefont {Ebbesen}},\ }\bibfield  {journal} {\bibinfo  {journal} {arXiv}\
  }\href {https://doi.org/10.48550/ARXIV.1911.01459}
  {10.48550/ARXIV.1911.01459} (\bibinfo {year} {2019})\BibitemShut {NoStop}%
\bibitem [{\citenamefont {Thomas}\ \emph {et~al.}(2021)\citenamefont {Thomas},
  \citenamefont {Devaux}, \citenamefont {Nagarajan}, \citenamefont {Rogez},
  \citenamefont {Seidel}, \citenamefont {Richard}, \citenamefont {Genet},
  \citenamefont {Drillon},\ and\ \citenamefont {Ebbesen}}]{FerroYBCO}%
  \BibitemOpen
  \bibfield  {author} {\bibinfo {author} {\bibfnamefont {A.}~\bibnamefont
  {Thomas}}, \bibinfo {author} {\bibfnamefont {E.}~\bibnamefont {Devaux}},
  \bibinfo {author} {\bibfnamefont {K.}~\bibnamefont {Nagarajan}}, \bibinfo
  {author} {\bibfnamefont {G.}~\bibnamefont {Rogez}}, \bibinfo {author}
  {\bibfnamefont {M.}~\bibnamefont {Seidel}}, \bibinfo {author} {\bibfnamefont
  {F.}~\bibnamefont {Richard}}, \bibinfo {author} {\bibfnamefont
  {C.}~\bibnamefont {Genet}}, \bibinfo {author} {\bibfnamefont
  {M.}~\bibnamefont {Drillon}},\ and\ \bibinfo {author} {\bibfnamefont {T.~W.}\
  \bibnamefont {Ebbesen}},\ }\href@noop {} {\bibfield  {journal} {\bibinfo
  {journal} {Nano Lett.}\ }\textbf {\bibinfo {volume} {21}},\ \bibinfo {pages}
  {4365} (\bibinfo {year} {2021})}\BibitemShut {NoStop}%
\bibitem [{\citenamefont {Garcia-Vidal}\ \emph {et~al.}(2019)\citenamefont
  {Garcia-Vidal}, \citenamefont {Ciuti},\ and\ \citenamefont
  {Ebbesen}}]{ScienceRev}%
  \BibitemOpen
  \bibfield  {author} {\bibinfo {author} {\bibfnamefont {F.~J.}\ \bibnamefont
  {Garcia-Vidal}}, \bibinfo {author} {\bibfnamefont {C.}~\bibnamefont
  {Ciuti}},\ and\ \bibinfo {author} {\bibfnamefont {T.~W.}\ \bibnamefont
  {Ebbesen}},\ }\href@noop {} {\bibfield  {journal} {\bibinfo  {journal}
  {Science}\ }\textbf {\bibinfo {volume} {373}},\ \bibinfo {pages} {eabd0336}
  (\bibinfo {year} {2019})}\BibitemShut {NoStop}%
\bibitem [{\citenamefont {Bednorz}\ and\ \citenamefont
  {M\"uller}(1986)}]{sconref1}%
  \BibitemOpen
  \bibfield  {author} {\bibinfo {author} {\bibfnamefont {J.~G.}\ \bibnamefont
  {Bednorz}}\ and\ \bibinfo {author} {\bibfnamefont {K.~A.}\ \bibnamefont
  {M\"uller}},\ }\href@noop {} {\bibfield  {journal} {\bibinfo  {journal} {Z.
  Phys. B - Condensed Matter}\ }\textbf {\bibinfo {volume} {64}},\ \bibinfo
  {pages} {189} (\bibinfo {year} {1986})}\BibitemShut {NoStop}%
\bibitem [{\citenamefont {Wu}\ \emph {et~al.}(1987)\citenamefont {Wu},
  \citenamefont {Ashburn}, \citenamefont {Torng}, \citenamefont {Hor},
  \citenamefont {Meng}, \citenamefont {Gao}, \citenamefont {Huang},
  \citenamefont {Wang},\ and\ \citenamefont {Chu}}]{sconref2}%
  \BibitemOpen
  \bibfield  {author} {\bibinfo {author} {\bibfnamefont {M.~K.}\ \bibnamefont
  {Wu}}, \bibinfo {author} {\bibfnamefont {J.~R.}\ \bibnamefont {Ashburn}},
  \bibinfo {author} {\bibfnamefont {C.~J.}\ \bibnamefont {Torng}}, \bibinfo
  {author} {\bibfnamefont {P.~H.}\ \bibnamefont {Hor}}, \bibinfo {author}
  {\bibfnamefont {R.~L.}\ \bibnamefont {Meng}}, \bibinfo {author}
  {\bibfnamefont {L.}~\bibnamefont {Gao}}, \bibinfo {author} {\bibfnamefont
  {Z.~J.}\ \bibnamefont {Huang}}, \bibinfo {author} {\bibfnamefont {Y.~Q.}\
  \bibnamefont {Wang}},\ and\ \bibinfo {author} {\bibfnamefont {C.~W.}\
  \bibnamefont {Chu}},\ }\href@noop {} {\bibfield  {journal} {\bibinfo
  {journal} {Phys. Rev. Lett.}\ }\textbf {\bibinfo {volume} {58}},\ \bibinfo
  {pages} {908} (\bibinfo {year} {1987})}\BibitemShut {NoStop}%
\bibitem [{\citenamefont {Hebard}\ \emph {et~al.}(1991)\citenamefont {Hebard},
  \citenamefont {Rosseinsky}, \citenamefont {Haddon}, \citenamefont {Murphy},
  \citenamefont {Glarum}, \citenamefont {Palstra}, \citenamefont {Ramirez},\
  and\ \citenamefont {Kortan}}]{Rb3C60scon1}%
  \BibitemOpen
  \bibfield  {author} {\bibinfo {author} {\bibfnamefont {A.~F.}\ \bibnamefont
  {Hebard}}, \bibinfo {author} {\bibfnamefont {M.~J.}\ \bibnamefont
  {Rosseinsky}}, \bibinfo {author} {\bibfnamefont {R.~C.}\ \bibnamefont
  {Haddon}}, \bibinfo {author} {\bibfnamefont {D.~W.}\ \bibnamefont {Murphy}},
  \bibinfo {author} {\bibfnamefont {S.~H.}\ \bibnamefont {Glarum}}, \bibinfo
  {author} {\bibfnamefont {T.~T.~M.}\ \bibnamefont {Palstra}}, \bibinfo
  {author} {\bibfnamefont {A.~P.}\ \bibnamefont {Ramirez}},\ and\ \bibinfo
  {author} {\bibfnamefont {A.~R.}\ \bibnamefont {Kortan}},\ }\href@noop {}
  {\bibfield  {journal} {\bibinfo  {journal} {Nature}\ }\textbf {\bibinfo
  {volume} {350}},\ \bibinfo {pages} {600} (\bibinfo {year}
  {1991})}\BibitemShut {NoStop}%
\bibitem [{\citenamefont {Holczer}\ \emph {et~al.}(1991)\citenamefont
  {Holczer}, \citenamefont {Klein}, \citenamefont {Huang}, \citenamefont
  {Kaner}, \citenamefont {Fu}, \citenamefont {Whetten},\ and\ \citenamefont
  {Diederich}}]{Rb3C60scon2}%
  \BibitemOpen
  \bibfield  {author} {\bibinfo {author} {\bibfnamefont {K.}~\bibnamefont
  {Holczer}}, \bibinfo {author} {\bibfnamefont {O.}~\bibnamefont {Klein}},
  \bibinfo {author} {\bibfnamefont {S.-M.}\ \bibnamefont {Huang}}, \bibinfo
  {author} {\bibfnamefont {R.~B.}\ \bibnamefont {Kaner}}, \bibinfo {author}
  {\bibfnamefont {K.-J.}\ \bibnamefont {Fu}}, \bibinfo {author} {\bibfnamefont
  {R.~L.}\ \bibnamefont {Whetten}},\ and\ \bibinfo {author} {\bibfnamefont
  {F.}~\bibnamefont {Diederich}},\ }\href@noop {} {\bibfield  {journal}
  {\bibinfo  {journal} {Science}\ }\textbf {\bibinfo {volume} {252}},\ \bibinfo
  {pages} {1154} (\bibinfo {year} {1991})}\BibitemShut {NoStop}%
\bibitem [{\citenamefont {Rosseinsky}\ \emph {et~al.}(1991)\citenamefont
  {Rosseinsky}, \citenamefont {Ramirez}, \citenamefont {Glarum}, \citenamefont
  {Murphy}, \citenamefont {Haddon}, \citenamefont {Hebard}, \citenamefont
  {Palstra}, \citenamefont {Kortan}, \citenamefont {Zahurak},\ and\
  \citenamefont {Makhija}}]{Rb3C60scon3}%
  \BibitemOpen
  \bibfield  {author} {\bibinfo {author} {\bibfnamefont {M.~J.}\ \bibnamefont
  {Rosseinsky}}, \bibinfo {author} {\bibfnamefont {A.~P.}\ \bibnamefont
  {Ramirez}}, \bibinfo {author} {\bibfnamefont {S.~H.}\ \bibnamefont {Glarum}},
  \bibinfo {author} {\bibfnamefont {D.~W.}\ \bibnamefont {Murphy}}, \bibinfo
  {author} {\bibfnamefont {R.~C.}\ \bibnamefont {Haddon}}, \bibinfo {author}
  {\bibfnamefont {A.~F.}\ \bibnamefont {Hebard}}, \bibinfo {author}
  {\bibfnamefont {T.~T.~M.}\ \bibnamefont {Palstra}}, \bibinfo {author}
  {\bibfnamefont {A.~R.}\ \bibnamefont {Kortan}}, \bibinfo {author}
  {\bibfnamefont {S.~M.}\ \bibnamefont {Zahurak}},\ and\ \bibinfo {author}
  {\bibfnamefont {A.~V.}\ \bibnamefont {Makhija}},\ }\href@noop {} {\bibfield
  {journal} {\bibinfo  {journal} {Phys. Rev. Lett.}\ }\textbf {\bibinfo
  {volume} {66}},\ \bibinfo {pages} {2830} (\bibinfo {year}
  {1991})}\BibitemShut {NoStop}%
\bibitem [{\citenamefont {Appugliese}\ \emph {et~al.}(2022)\citenamefont
  {Appugliese}, \citenamefont {Enkner}, \citenamefont {Paravicini-Bagliani},
  \citenamefont {Beck}, \citenamefont {Reichl}, \citenamefont {Wegscheider},
  \citenamefont {Scalari}, \citenamefont {Ciuti},\ and\ \citenamefont
  {Faist}}]{Hallpaper}%
  \BibitemOpen
  \bibfield  {author} {\bibinfo {author} {\bibfnamefont {F.}~\bibnamefont
  {Appugliese}}, \bibinfo {author} {\bibfnamefont {J.}~\bibnamefont {Enkner}},
  \bibinfo {author} {\bibfnamefont {G.~L.}\ \bibnamefont
  {Paravicini-Bagliani}}, \bibinfo {author} {\bibfnamefont {M.}~\bibnamefont
  {Beck}}, \bibinfo {author} {\bibfnamefont {C.}~\bibnamefont {Reichl}},
  \bibinfo {author} {\bibfnamefont {W.}~\bibnamefont {Wegscheider}}, \bibinfo
  {author} {\bibfnamefont {G.}~\bibnamefont {Scalari}}, \bibinfo {author}
  {\bibfnamefont {C.}~\bibnamefont {Ciuti}},\ and\ \bibinfo {author}
  {\bibfnamefont {J.}~\bibnamefont {Faist}},\ }\href@noop {} {\bibfield
  {journal} {\bibinfo  {journal} {Science}\ }\textbf {\bibinfo {volume}
  {375}},\ \bibinfo {pages} {1030} (\bibinfo {year} {2022})}\BibitemShut
  {NoStop}%
\bibitem [{\citenamefont {Mitrano}\ \emph {et~al.}(2016)\citenamefont
  {Mitrano}, \citenamefont {Cantaluppi}, \citenamefont {Nicoletti},
  \citenamefont {Kaiser}, \citenamefont {Perucchi}, \citenamefont {Lupi},
  \citenamefont {Pietro}, \citenamefont {Pontiroli}, \citenamefont {Riccò},
  \citenamefont {S.~R. Clar~and},\ and\ \citenamefont {Cavalleri}}]{pumppaper}%
  \BibitemOpen
  \bibfield  {author} {\bibinfo {author} {\bibfnamefont {M.}~\bibnamefont
  {Mitrano}}, \bibinfo {author} {\bibfnamefont {A.}~\bibnamefont {Cantaluppi}},
  \bibinfo {author} {\bibfnamefont {D.}~\bibnamefont {Nicoletti}}, \bibinfo
  {author} {\bibfnamefont {S.}~\bibnamefont {Kaiser}}, \bibinfo {author}
  {\bibfnamefont {A.}~\bibnamefont {Perucchi}}, \bibinfo {author}
  {\bibfnamefont {S.}~\bibnamefont {Lupi}}, \bibinfo {author} {\bibfnamefont
  {P.~D.}\ \bibnamefont {Pietro}}, \bibinfo {author} {\bibfnamefont
  {D.}~\bibnamefont {Pontiroli}}, \bibinfo {author} {\bibfnamefont
  {M.}~\bibnamefont {Riccò}}, \bibinfo {author} {\bibfnamefont {D.~J.}\
  \bibnamefont {S.~R. Clar~and}},\ and\ \bibinfo {author} {\bibfnamefont
  {A.}~\bibnamefont {Cavalleri}},\ }\href@noop {} {\bibfield  {journal}
  {\bibinfo  {journal} {Nature doi:10.1038/nature16522}\ }\textbf {\bibinfo
  {volume} {530}},\ \bibinfo {pages} {461} (\bibinfo {year}
  {2016})}\BibitemShut {NoStop}%
\bibitem [{\citenamefont {Chu}\ and\ \citenamefont {McHenry}(1997)}]{McHenry}%
  \BibitemOpen
  \bibfield  {author} {\bibinfo {author} {\bibfnamefont {S.}~\bibnamefont
  {Chu}}\ and\ \bibinfo {author} {\bibfnamefont {M.~E.}\ \bibnamefont
  {McHenry}},\ }\href@noop {} {\bibfield  {journal} {\bibinfo  {journal} {Phys.
  Rev. B}\ }\textbf {\bibinfo {volume} {55}},\ \bibinfo {pages} {11722}
  (\bibinfo {year} {1997})}\BibitemShut {NoStop}%
\bibitem [{\citenamefont {Hurd}(1965)}]{Hurd1963}%
  \BibitemOpen
  \bibfield  {author} {\bibinfo {author} {\bibfnamefont {C.~M.}\ \bibnamefont
  {Hurd}},\ }\href@noop {} {\bibfield  {journal} {\bibinfo  {journal} {Rev.
  Sci. Instrum.}\ }\textbf {\bibinfo {volume} {37}},\ \bibinfo {pages} {515}
  (\bibinfo {year} {1965})}\BibitemShut {NoStop}%
\bibitem [{\citenamefont {Freiman}\ and\ \citenamefont
  {Jodl}(2004)}]{Freiman2004}%
  \BibitemOpen
  \bibfield  {author} {\bibinfo {author} {\bibfnamefont {Y.~A.}\ \bibnamefont
  {Freiman}}\ and\ \bibinfo {author} {\bibfnamefont {H.~J.}\ \bibnamefont
  {Jodl}},\ }\href@noop {} {\bibfield  {journal} {\bibinfo  {journal} {Phys.
  Reports}\ }\textbf {\bibinfo {volume} {401}},\ \bibinfo {pages} {1} (\bibinfo
  {year} {2004})}\BibitemShut {NoStop}%
\bibitem [{\citenamefont {Giannetta}\ \emph {et~al.}(2022)\citenamefont
  {Giannetta}, \citenamefont {Carrington},\ and\ \citenamefont
  {Prozorov}}]{Gianetta2022}%
  \BibitemOpen
  \bibfield  {author} {\bibinfo {author} {\bibfnamefont {R.}~\bibnamefont
  {Giannetta}}, \bibinfo {author} {\bibfnamefont {A.}~\bibnamefont
  {Carrington}},\ and\ \bibinfo {author} {\bibfnamefont {R.}~\bibnamefont
  {Prozorov}},\ }\href@noop {} {\bibfield  {journal} {\bibinfo  {journal}
  {Jour. Low Temp. Phys.}\ }\textbf {\bibinfo {volume} {208}},\ \bibinfo
  {pages} {119} (\bibinfo {year} {2022})}\BibitemShut {NoStop}%
\bibitem [{\citenamefont {Chauvet}\ \emph {et~al.}(1994)\citenamefont
  {Chauvet}, \citenamefont {Oszl\'anyi}, \citenamefont {Forro}, \citenamefont
  {Stephens}, \citenamefont {Tegze}, \citenamefont {Faigel},\ and\
  \citenamefont {J\'anossy}}]{RbC60SDW1994PRL}%
  \BibitemOpen
  \bibfield  {author} {\bibinfo {author} {\bibfnamefont {O.}~\bibnamefont
  {Chauvet}}, \bibinfo {author} {\bibfnamefont {G.}~\bibnamefont {Oszl\'anyi}},
  \bibinfo {author} {\bibfnamefont {L.}~\bibnamefont {Forro}}, \bibinfo
  {author} {\bibfnamefont {P.~W.}\ \bibnamefont {Stephens}}, \bibinfo {author}
  {\bibfnamefont {M.}~\bibnamefont {Tegze}}, \bibinfo {author} {\bibfnamefont
  {G.}~\bibnamefont {Faigel}},\ and\ \bibinfo {author} {\bibfnamefont
  {A.}~\bibnamefont {J\'anossy}},\ }\href@noop {} {\bibfield  {journal}
  {\bibinfo  {journal} {Phys. Rev. Lett.}\ }\textbf {\bibinfo {volume} {72}},\
  \bibinfo {pages} {2721} (\bibinfo {year} {1994})}\BibitemShut {NoStop}%
\bibitem [{\citenamefont {Diederichs}\ \emph {et~al.}(1997)\citenamefont
  {Diederichs}, \citenamefont {Schilling}, \citenamefont {Herwig},\ and\
  \citenamefont {Yelon}}]{Schilling}%
  \BibitemOpen
  \bibfield  {author} {\bibinfo {author} {\bibfnamefont {J.}~\bibnamefont
  {Diederichs}}, \bibinfo {author} {\bibfnamefont {J.~S.}\ \bibnamefont
  {Schilling}}, \bibinfo {author} {\bibfnamefont {K.~W.}\ \bibnamefont
  {Herwig}},\ and\ \bibinfo {author} {\bibfnamefont {W.~B.}\ \bibnamefont
  {Yelon}},\ }\href@noop {} {\bibfield  {journal} {\bibinfo  {journal} {J.
  Phys. Chem. Solids}\ }\textbf {\bibinfo {volume} {58}},\ \bibinfo {pages}
  {123} (\bibinfo {year} {1997})}\BibitemShut {NoStop}%
\bibitem [{\citenamefont {Garcia}\ \emph {et~al.}(2009)\citenamefont {Garcia},
  \citenamefont {Pinel}, \citenamefont {{de la Venta}}, \citenamefont
  {Quesada}, \citenamefont {Bouzas}, \citenamefont {Fern\'andez}, \citenamefont
  {Romero}, \citenamefont {Gonz\'alez},\ and\ \citenamefont
  {Costa-Kr\"amer}}]{Garcia2009}%
  \BibitemOpen
  \bibfield  {author} {\bibinfo {author} {\bibfnamefont {M.~A.}\ \bibnamefont
  {Garcia}}, \bibinfo {author} {\bibfnamefont {E.~F.}\ \bibnamefont {Pinel}},
  \bibinfo {author} {\bibfnamefont {J.}~\bibnamefont {{de la Venta}}}, \bibinfo
  {author} {\bibfnamefont {A.}~\bibnamefont {Quesada}}, \bibinfo {author}
  {\bibfnamefont {V.}~\bibnamefont {Bouzas}}, \bibinfo {author} {\bibfnamefont
  {J.~F.}\ \bibnamefont {Fern\'andez}}, \bibinfo {author} {\bibfnamefont
  {J.~J.}\ \bibnamefont {Romero}}, \bibinfo {author} {\bibfnamefont {M.~S.~M.}\
  \bibnamefont {Gonz\'alez}},\ and\ \bibinfo {author} {\bibfnamefont {J.~L.}\
  \bibnamefont {Costa-Kr\"amer}},\ }\href@noop {} {\bibfield  {journal}
  {\bibinfo  {journal} {Jour. Appl. Phys.}\ }\textbf {\bibinfo {volume}
  {105}},\ \bibinfo {pages} {013925} (\bibinfo {year} {2009})}\BibitemShut
  {NoStop}%
\bibitem [{\citenamefont {Panagopoulos}\ \emph {et~al.}(1996)\citenamefont
  {Panagopoulos}, \citenamefont {Zhou}, \citenamefont {Athanassopoulou},\ and\
  \citenamefont {Cooper}}]{Panagopoulos1996}%
  \BibitemOpen
  \bibfield  {author} {\bibinfo {author} {\bibfnamefont {C.}~\bibnamefont
  {Panagopoulos}}, \bibinfo {author} {\bibfnamefont {W.}~\bibnamefont {Zhou}},
  \bibinfo {author} {\bibfnamefont {N.}~\bibnamefont {Athanassopoulou}},\ and\
  \bibinfo {author} {\bibfnamefont {J.~R.}\ \bibnamefont {Cooper}},\
  }\href@noop {} {\bibfield  {journal} {\bibinfo  {journal} {Phys. C:
  Superconductivity}\ }\textbf {\bibinfo {volume} {269}},\ \bibinfo {pages}
  {157} (\bibinfo {year} {1996})}\BibitemShut {NoStop}%
\bibitem [{\citenamefont {Guskos}\ \emph {et~al.}(1995)\citenamefont {Guskos},
  \citenamefont {Likodimos}, \citenamefont {Londos}, \citenamefont {Psycharis},
  \citenamefont {Mitros}, \citenamefont {Koufoudakis}, \citenamefont
  {Gamari-Seale}, \citenamefont {Windsch},\ and\ \citenamefont
  {Metz}}]{Guskos1995}%
  \BibitemOpen
  \bibfield  {author} {\bibinfo {author} {\bibfnamefont {N.}~\bibnamefont
  {Guskos}}, \bibinfo {author} {\bibfnamefont {V.}~\bibnamefont {Likodimos}},
  \bibinfo {author} {\bibfnamefont {C.}~\bibnamefont {Londos}}, \bibinfo
  {author} {\bibfnamefont {V.}~\bibnamefont {Psycharis}}, \bibinfo {author}
  {\bibfnamefont {C.}~\bibnamefont {Mitros}}, \bibinfo {author} {\bibfnamefont
  {A.}~\bibnamefont {Koufoudakis}}, \bibinfo {author} {\bibfnamefont
  {H.}~\bibnamefont {Gamari-Seale}}, \bibinfo {author} {\bibfnamefont
  {W.}~\bibnamefont {Windsch}},\ and\ \bibinfo {author} {\bibfnamefont
  {H.}~\bibnamefont {Metz}},\ }\href@noop {} {\bibfield  {journal} {\bibinfo
  {journal} {Journal Solid State Chemistry}\ }\textbf {\bibinfo {volume}
  {119}},\ \bibinfo {pages} {50} (\bibinfo {year} {1995})}\BibitemShut
  {NoStop}%
\bibitem [{\citenamefont {Aranda}\ and\ \citenamefont
  {Attfield}(1993)}]{Aranda}%
  \BibitemOpen
  \bibfield  {author} {\bibinfo {author} {\bibfnamefont {M.~A.~G.}\
  \bibnamefont {Aranda}}\ and\ \bibinfo {author} {\bibfnamefont {J.~P.}\
  \bibnamefont {Attfield}},\ }\href@noop {} {\bibfield  {journal} {\bibinfo
  {journal} {Angew, Chem. Int. Ed. Eng.}\ }\textbf {\bibinfo {volume} {32}},\
  \bibinfo {pages} {1454} (\bibinfo {year} {1993})}\BibitemShut {NoStop}%
\bibitem [{\citenamefont {Biscaras}\ \emph {et~al.}(2012)\citenamefont
  {Biscaras}, \citenamefont {Leridon}, \citenamefont {Colson}, \citenamefont
  {Forget},\ and\ \citenamefont {Monod}}]{Monod2012}%
  \BibitemOpen
  \bibfield  {author} {\bibinfo {author} {\bibfnamefont {J.}~\bibnamefont
  {Biscaras}}, \bibinfo {author} {\bibfnamefont {B.}~\bibnamefont {Leridon}},
  \bibinfo {author} {\bibfnamefont {D.}~\bibnamefont {Colson}}, \bibinfo
  {author} {\bibfnamefont {A.}~\bibnamefont {Forget}},\ and\ \bibinfo {author}
  {\bibfnamefont {P.}~\bibnamefont {Monod}},\ }\href@noop {} {\bibfield
  {journal} {\bibinfo  {journal} {Phys. Rev. B}\ }\textbf {\bibinfo {volume}
  {85}},\ \bibinfo {pages} {134517} (\bibinfo {year} {2012})}\BibitemShut
  {NoStop}%
\bibitem [{\citenamefont {Kokanovi\ifmmode~\acute{c}\else \'{c}\fi{}}\ and\
  \citenamefont {Cooper}(2016)}]{Kokanovic2016}%
  \BibitemOpen
  \bibfield  {author} {\bibinfo {author} {\bibfnamefont {I.}~\bibnamefont
  {Kokanovi\ifmmode~\acute{c}\else \'{c}\fi{}}}\ and\ \bibinfo {author}
  {\bibfnamefont {J.~R.}\ \bibnamefont {Cooper}},\ }\href@noop {} {\bibfield
  {journal} {\bibinfo  {journal} {Phys. Rev. B}\ }\textbf {\bibinfo {volume}
  {94}},\ \bibinfo {pages} {075155} (\bibinfo {year} {2016})}\BibitemShut
  {NoStop}%
\end{thebibliography}%

\end{document}